\documentclass{aastex}
\usepackage{spr-astr-addons}

\begin{document}

\title{Scenario of Accelerating Universe from the Phenomenological $\Lambda$ Models}

\shorttitle{Accelerating Universe}

\shortauthors{Ray et al.}

\author{Saibal Ray\altaffilmark{1}}
\altaffiltext{1}{Department of Physics, Government College of
Engineering and Ceramic Technology, Kolkata - 700 010, West
Bengal, India \\ e-mail: saibal@iucaa.ernet.in}

\and

\author{Utpal Mukhopadhyay\altaffilmark{2}}
\altaffiltext{2}{Satyabharati Vidyapith, Nabapalli, North 24
Parganas, Kolkata - 700 126, West Bengal, India\\e-mail:
utpalsbv@gmail.com}

\and

\author{Farook Rahaman\altaffilmark{3}}
\altaffiltext{3}{Department of Mathematics, Jadavpur University,
Kolkata - 700 032, West Bengal, India\\e-mail:
farook@iucaa.ernet.in}

\and

\author{Ruby Sarkar\altaffilmark{4}}

\altaffiltext{4}{Department of Mathematics, Jadavpur University,
Kolkata - 700 032, West Bengal, India\\e-mail:
rubyfantacy@yahoo.co.in}

\begin{abstract}
Dark matter, the major component of the matter content of the
Universe, played a significant role   at early stages during
structure formation. But at present the Universe is dark energy
dominated as well as accelerating. Here, the presence of dark
energy has been established by including a time-dependent
$\Lambda$ term in the Einstein's field equations. This model is
compatible with the idea of an accelerating Universe so far as the
value of the deceleration parameter is concerned. Possibility of a
change in sign of the deceleration parameter is also discussed.
The impact of considering the speed of light as variable in the
field equations has also been investigated by using a well known
time-dependent $\Lambda$ model.
\end{abstract}


\section{Introduction}
Dark matter and dark energy are two major constituents of the
present Universe. As it stands today, the Universe is composed of
nearly $30\%$ matter and $70\%$ dark energy \citep{Kirshner2003}.
Again, of the total matter content, about $25\%$ are non-luminous
or dark while baryonic matter contributes with the $5\%$ of all
energetic content of the Universe and surprisingly its largest
fraction is found in the Intergalactic Medium, not in the galaxies
\citep{Ostriker2003,Shull2005,Ettori2009}. Long ago, using Virial
Theorem, \citet{Zwicky1937} for the first time suggested about the
existence of dark matter. Afterwards, galactic rotation curve
studies also supported Zwicky's idea
\citep{Roberts1973,Ostriker1974,Einasto1974,Rubin1978}. When, as a
consequence of Inflationary Theory of \citet{Guth1981}
and~others~\citep{Linde1982,Albrecht1982}, it became clear that
the Universe must be flat, cosmologists became convinced that
$96\%$ matter content of the Universe should be hidden mass. But,
this simplified cosmological picture soon ran into trouble since,
in spite of an intense search, evidence in favor of such a huge
amount of dark matter was lacking. Theoretical scientists then
speculated that matter energy density of the Universe cannot be
more than one-third of the total energy density and hence the
remaining two-third energy density should be compensated by a
cosmological constant \citep{Turner2003}. Finally, observational
result for an accelerating Universe
\citep{Perlmutter1998,Riess1998} favored the above speculation and
the idea of an accelerating agent, termed as dark energy, was
accepted.

Both dark matter and dark energy have played their specific roles
in different stages of cosmic evolution. Dark matter had a
significant role in the early Universe during structure formation
because its nature is to clump in sub-megaparsec scales. Although
the exact composition of dark matter is still unknown, COBE and
CMB experiments suggest that baryonic dark matter cannot be more
than a small fraction of total dark matter present in the Universe
\citep{Sahni2004}. Moreover, observational constraint regarding
neutrino mass and relic neutrino density
\citep{Minakata2002,Elgaroy2002,Spergel2003,Ellis2003} eliminate
the possibility of hot dark matter. So, cold non-baryonic dark
matter which can clump on small scales is favoured now
\citep{Sahni2004}. The Standard Cold Dark Matter (SCDM) model,
introduced in the early 1980's, is presently disfavoured and it is
replaced by $\Lambda$-CDM model in the context of an accelerating
Universe. $\Lambda$-CDM model is found to be in nice agreement
with various observational results \citep{Tegmark2003} and as an
advantage of assuming a nearly scale-invariant primordial
perturbations and a Universe with no spatial curvature as
predicted by the Inflationary theory
\citep{Mukhanov1981,Guth1982,Hawking1982,Starobinsky1982,Bardeen1982}.
But, in the $\Lambda$-CDM scenario the present acceleration of the
Universe cannot be a permanent feature because, structure
formation cannot proceed during acceleration. In fact, some recent
works \citep{Padmanabhan2002,Amendola2003} show that the present
accelerating phase was preceded by a decelerating one and
observational evidence \citep{Riess2001} also supports this idea.
So, the deceleration parameter must have undergone a flip in sign
during cosmic evolution.

Let us now move towards the cosmological constant problem. There
are actually two fine-tuning problems with cosmological constant:
(i) the value of $\Lambda$ must be 123 orders of magnitude and 55
orders of magnitude larger on the Planck scale ($T \sim 10^{19}$
GeV) and the electroweak scale ($T \sim 10^2$ GeV), respectively,
than its presently observed value, and (ii) the matter and
radiation energy densities of the expanding Universe fall off as
$a^{-3}$ and $a^{-4}$, respectively, where $a$ is the scale factor
of the universe, while $\Lambda$ remains constant. The only
solutions to these problems is to assume a dynamical character of
constant $\Lambda$, especially a time-dependent $\Lambda$ which
has decreased slowly from its large initial value to reach its
present small value \citep{Overduin1998}. This idea of
time-varying cosmological constant $\Lambda$ gives us a motivation
behind the present investigation.

Various dark energy models have been proposed during the last
decade or so (for an overview see the works of
\citet{Overduin1998} and \citet{Sahni2000}). One of the favorite
candidates among these, obviously, are the models related to
dynamic $\Lambda$ term. In a recent work \citep{Ray2007a},
equivalence of three dynamical $\Lambda$ models viz.
$\Lambda\simeq H^2$,$\Lambda\simeq {\ddot a/a}$ and $\Lambda\simeq
\rho$ has been established and in another work \citep{Ray2007b}
age of the Universe is calculated using the same three $\Lambda$
models. In this work, without taking recourse of any specific
model, an overall study of accelerating Universe is done.
Importance of $\Lambda$ for an accelerating Universe is also
revealed with special reference to the work of \citet{Deb1999}.

Although a large number of dark energy models with both constant
and variable $\Lambda$ are found in the literature, the very
phrase `varying speed of light' itself, in general, has a
shuddering effect on mind owing to the panic for collapse of the
grand edifice of modern physics through the breakdown of special
and general theory of relativity. This situation can be compared
with the idea of variation of the gravitational constant $G$ in
the pre-relativistic era when Newton's law of Universal
Gravitation was the most precious jewel in the kingdom of
classical physics. The situation has been reversed after $1905$
and at present any idea about a possible change in the speed of
light $c$ is considered as a mark of iconoclastic attitude.
However, \citet{Michell1784} showed that a particular mass to
radius ratio of a star implies that the escape velocity of the
star would be equal to the speed of light. But the earliest
inception of the idea of changing speed of light before the advent
of special theory of relativity is due to ~\citet{Thomson1874}.
The relativistic era has seen many ups and downs regarding the
possibility of variation of $c$. Although \citet{Einstein1911} and
\citet{Feynman1988} themselves were not afraid of thinking in
terms of a changing speed of light, \citet{Eddington1946}
vehemently opposed the idea of a variable $c$ by saying ``A
variation of $c$ is self-contradictory''. In the 1930s, variation
of $c$ came into limelight for providing an alternative
explanation of cosmological
redshift~\cite{Stewart1931,Buc1932,Wold1935}.

But, recent varying speed of light (VSL) theories have a basic
difference with the previous ones in the sense that most of them
are entangled with the hot Big Bang model of the Universe. The
first seminal paper with this cosmological background is due to
\citet{Moffat1993}. In that \citep{Moffat1993} and in a subsequent
paper \citep{Moffat2002}, Moffat has presented his VSL theory by
invoking the idea of phase transition which can solve the horizon
problem without taking recourse of the popular idea of inflation.
A major problem with the variation of light speed is the violation
of energy-conservation law of special theory of relativity, viz.
$E = mc^2$, where $c$ is a constant although alternative
derivations of this relation without making relativistic idea was
shown to be possible~\citep{Poincare1900,Born1962}. But, this
problem can be avoided if in Einstein field equations

\begin{eqnarray} G_{ij} = \kappa T_{ij}, \end{eqnarray}

(where as weak field approximation $\kappa$ comes out as $8\pi
G/c^4$) only the energy component of $T_{ij}$ is taken as $mc^2$
then equation (1) comes out as

\begin{eqnarray} G_{ij} = \frac{8\pi Gm}{c^2}.\end{eqnarray}

\noindent The dimensionality of above equation (2) can be
accounted for through the component {\it time-time} of the
Einstein field equation for the tensor energy-momentum of fluid
perfect. It is clear from this equation (2) that $c$ is inversely
proportional to the curvature of space-time and for any departure
from $\Omega\neq 1$, the speed of light and the curvature term
will adjust themselves so as to make the Universe flat. So, from
this point of view, the variation in the speed of light is
essential.

After observational evidence in favour of the redshift dependence
of the fine structure constant $\alpha$ is
established~\citep{Web1999,Web2001,Murphy2001,Web2003}, VSL
theories from various point of view has come up. Since $\alpha$ is
related to the velocity of light $c$ through the relation $\alpha
= e^2/hc$ (where $e$ is the charge of a proton and $h$ is Planck's
constant), variability of $\alpha$ implies variability of $c$
provided constancy of $e$ and $h$ is assumed. In dilaton theories,
variation of $c$ is assumed
~\citep{Bekenstein1982,Barrow2001,Olive2001,Sandvik2002,Bekenstein2002,Martins2002,Uzan2003},
but VSL theories pinpoint on the variation of $c$
~\citep{Peres1967,Barrow1998,Barrow1999,Albrecht1999,Barrow2000,Moffat2002,Peres2002}
although in some cases~\citep{Magueijo2000} $h$ is thought to be
responsible for the variability of $\alpha$. So, it is quite
natural to investigate the possible variation of $c$ in the
context of the present accelerating Universe, discovered through
SN Ia observations~\citep{Perlmutter1998,Riess1998}. In one such
work, \citet{Camare2007} have investigated the behaviour of two
time-varying models of $c$, viz. $c(t)\propto a^{-r}$ and
$c(t)\propto H^u$ where $a$ is the scale factor and $H$ is the
Hubble parameter. The present work is motivated by an intention to
investigate analytically the behaviour of a time-varying model
of~$c$.

In the first portion of the present investigation, without taking
recourse of any specific model, an overall study of accelerating
Universe has been done. Importance of dynamic $\Lambda$ for an
accelerating Universe is also revealed with special reference to
the work of \citet{Deb1999}. In the second part, the question of
constancy and variability of the speed of light has been dealt in
the framework of the present accelerating Universe by including a
time-dependent $\Lambda$ model in the field equations. In various
subsections of Sections 2 and 3, the role of $\Lambda$ has been
demonstrated without resorting to any specific $\Lambda$ model. In
Secs. 4 and 5, respectively Einstein's field equations for
variable $c$ have been solved analytically by choosing a
particular time-dependent model of $\Lambda$ and some salient
features of the solution have been discussed. Finally, some
specific conclusions arrived at from the present work have been
presented in Sec. 6.

\section{Field Equations with Constant Speed of Light}
We know that Einstein's field equation (including cosmological
parameter $\Lambda$) are given by
\begin{eqnarray}
  R^{ij} - \frac{1}{2}Rg^{ij} = -8\pi G\left[T^{ij} - \frac{\Lambda}{8\pi
  G}g^{ij}\right],
\end{eqnarray}
where the speed of light $c=1$ in relativistic unit and hence is a
constant quantity. For the spherically symmetric FLRW metric the
above equation yield respectively Friedmann equation and
Raychaudhuri equation given by
 \begin{eqnarray}
\left(\frac{\dot a}{a}\right)^{2}+\frac{k}{a^2}=\frac{8\pi
G\rho}{3}+\frac{\Lambda}{3},
\end{eqnarray}
\begin{eqnarray}
\left(\frac{\ddot a}{a}\right)=-\frac{4\pi
G}{3}(\rho+3p)+\frac{\Lambda}{3}
 \end{eqnarray}
where $\Lambda=\Lambda(t)$ is time-dependent, $k$ is the curvature
constant and $a(t)$ is the scale factor of the Universe.

From the equation (4) we have
\begin{eqnarray}
\rho=\frac{3}{8\pi
G}\left(\frac{k}{a^2}+H^2-\frac{\Lambda}{3}\right).
 \end{eqnarray}
So, the present energy density is given by
\begin{eqnarray}
\rho_0=\frac{3}{8\pi
G}\left(\frac{k}{a_0^2}+H_0^2-\frac{\Lambda_0}{3}\right)
\end{eqnarray}
where the suffix zero indicates the present values of the
corresponding cosmological parameters. Again, the deceleration
parameter $q$ is given by
\begin{eqnarray}
q=-\frac{a \ddot a}{\dot a^2}=-\frac{1}{H^2}\left(\frac{\ddot
a}{a}\right).
 \end{eqnarray}
Therefore,
\begin{eqnarray}
\frac{\ddot a}{a}=-qH^2.
 \end{eqnarray}
Using the equations (9) and (6), we get from (5),
\begin{eqnarray}
-qH^2=-\frac{1}{2}\left(\frac{k}{a^2}+H^2-\frac{\Lambda}{3}\right)-4\pi
Gp+\frac{\Lambda}{3}
\end{eqnarray}
which, after simplification, provides the expression for pressure
given by
\begin{eqnarray}
p=-\frac{1}{8\pi G}\left[\frac{k}{a^2}+(1-2q)H^2-\Lambda\right].
\end{eqnarray}
Thus, the present value for the pressure is given by
\begin{eqnarray}
p_0=-\frac{1}{8\pi
G}\left[\frac{k}{a_0^2}+(1-2q_0)H_0^2-\Lambda_0\right].
\end{eqnarray}
Let us choose the barotropic equation of state
\begin{eqnarray}
p=\omega\rho
\end{eqnarray}
where $\omega$ is the barotropic index or equation of state
parameter such that $\omega=p/\rho$. In general, $\omega$ is a
function of time, scale factor or redshift but sometimes it is
convenient to consider $\omega$ as a constant quantity because
current observational data has limited power to distinguish
between a time varying and constant equation of state
\citep{Kujat2002,Bartelmann2005}. However, we shall see in Sec.
3.3 that assumption of $\omega$ as constant in time will provide
interesting physical scenario.

For flat ($k=0$) Universe, using equations (6) and (11) in (13),
we get
\begin{eqnarray}
\omega=\frac{3\Omega_{\Lambda}+2q-1}{3(1-\Omega_{\Lambda})}
\end{eqnarray}
where $\Omega_{\Lambda}=\Lambda/3H^2$ is the vacuum energy density
of the Universe.

\begin{figure}
\begin{center}
\vspace{0.5cm}
\includegraphics[width=0.5\textwidth]{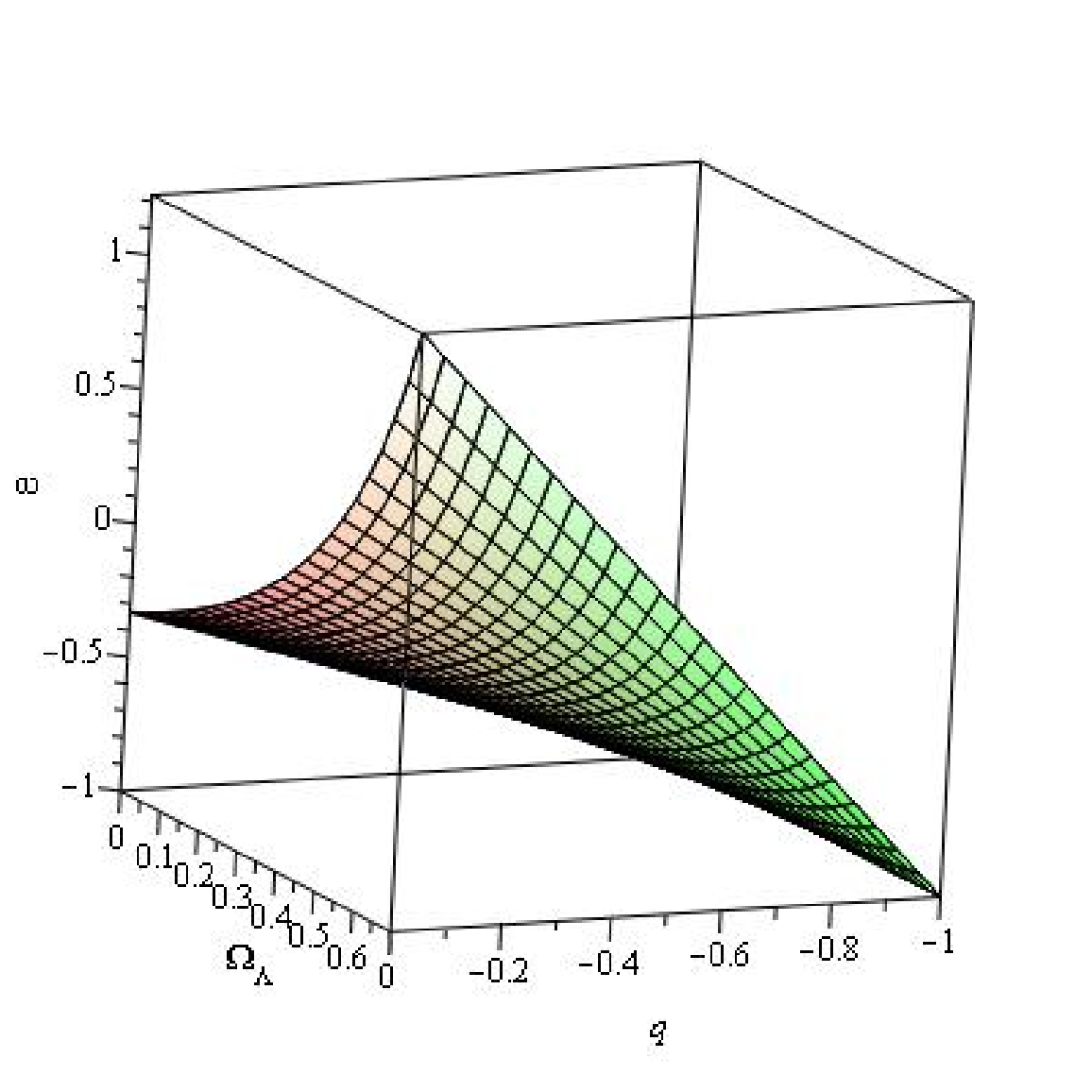}
\caption{The variation of equation of state parameter $\omega$
with respect to the vacuum energy density of the Universe ($
\Omega_\Lambda $) and the deceleration parameter ($q$) of the
Universe.}
\end{center}
\end{figure}

Similarly, for flat Universe, equations (7) and (12) can
respectively be written as
\begin{eqnarray}
\rho_0=\frac{3H_0^2}{8\pi G}(1-\Omega_{\Lambda_0})
\end{eqnarray}
and
\begin{eqnarray}
p_0=-\frac{H_0^2}{8\pi G}[(1-2q_0)-3\Omega_{\Lambda_0}]
\end{eqnarray}
where $\Omega_{\Lambda_0}$ is the present value of the vacuum
energy density.

From equations (14), (15) and (16) it is clear that the fate of
the Universe depends on $q$, $\Omega_{\Lambda}$ and $H$ (Fig. 1).
Also it is clear from equation (13) that for physical reality
$\Omega _{\Lambda_0}$ must be less than one. We also know that the
case $p_0>0$ provides a collapsing Universe. So, equation (16)
tells us that for a collapsing Universe, we must have
$(1-2q_0)<3\Omega _{\Lambda_0}$.

\section{Physical Features of the Model}

\subsection{Calculation of $\rho_c$, $\rho_0$ and $\rho_G$}
An important quantity which determines the future of the Universe
is the critical density $\rho_c$. The Universe is open or closed
according as the present density $\rho_0$ of the Universe is less
or greater than the critical density. Since at present, the
accepted value of the Hubble parameter $H_0$ is $(72 \pm 8)~km
s^{-1} Mpc^{-1}$ \citep{Kirshner2003}, we may choose $H_0=72~km
s^{-1} Mpc^{-1}$ for our calculation. For this value of $H_0$, we
get $\rho_c=3H_0^2/8\pi G \sim 9\times 10^{-30}~gcm^{-3}$.

In one of our previous work \citep{Ray2007a} we have shown that
the three kinematical $\Lambda$ models, viz. $\Lambda \sim (\dot
a/a)^2$, $\Lambda \sim (\ddot a/a)$ and $\Lambda \sim \rho$ are
equivalent and for  these models, $\rho_0=3 \times
10^{-30}~gcm^{-3}$. Also, the measured galactic mass density
$\rho_G$ is given by~\citet{Deb1999} $\rho_G=3.1\times
10^{-31}~gcm^{-3}$. Therefore,
\begin{eqnarray}
\frac{\rho_G}{\rho_c} \sim 0.033.
\end{eqnarray}
Also,
\begin{eqnarray}
\frac{\rho_0}{\rho_c}=\frac{1}{3}.
\end{eqnarray}
From equations (17) and (18) we, immediately, obtain
\begin{eqnarray}
\rho_G \sim 0.1\rho_0.
\end{eqnarray}
It is clear from equation (19) that galactic mass density is about
$10\%$ of the total mass density  of the present Universe. Hence,
there must be some hidden mass. Also, equation (18) implies that
the present total density of the Universe is one-third of the
critical density. This means that galactic (luminous) mass-density
is about $3\%$ of the critical density.

Again, the equation (7) can be written as
\begin{eqnarray}
\frac{k}{a^2}=\frac{1}{3}\left[{\Lambda_0}-8\pi G
(\rho_c-\rho_0)\right].
\end{eqnarray}
From equation (18) it is easy to see that $\rho_0<\rho_c$ and
hence $(\rho_c-\rho_0)>0$. Also, present observational results
indicate that, the Universe is flat (k=0). So, for a flat Universe
we must have
\begin{eqnarray}
\Lambda_0=8\pi G(\rho_c-\rho_0).
\end{eqnarray}
On the other hand, for a closed Universe, $\Lambda_0>8\pi G
(\rho_c-\rho_0)$ whereas for an open Universe, $\Lambda_0<8\pi G
(\rho_c-\rho_0)$. So, the cosmological parameter is an important
factor for determining the geometry of the Universe. Now, one of
the predictions of the inflationary theory is a flat Universe and
$\Lambda$ had a large value in the early stages of the Universe.
So, one may argue that it is the cosmological parameter which made
the Universe flat during inflation.

\subsection{Calculation of $q_0$}

For pressureless non-relativistic matter, $p=0$. Then from
equation (12) we have,
\begin{eqnarray}
\frac{k}{a^2}=\Lambda_0-(1-2q_0)H_0^2.
\end{eqnarray}
Using equation (22) in (7), we get
\begin{eqnarray}
\rho_0=\frac{1}{4\pi G}(\Lambda_0+3q_0 H_0^2).
\end{eqnarray}
Therefore,
\begin{eqnarray}
\frac{\rho_0}{\rho_c}=2\left(q_0+\frac{\Lambda_0}{3H_0^2}\right)=2(q_0+\Omega_{\Lambda_0}).
\end{eqnarray}
Using equation (18) and noting that $\Omega_{\Lambda_ 0}$ is
nearly equal to $0.7$ we get from equation (22) that $q_0$ is
about $-0.53$. This value of $q_0$ is in excellent agreement with
the present accepted value of this parameter \citep{Sahni2000} and
represents an accelerating Universe. Moreover, \citet{Deb1999},
without taking into account the cosmological parameter $\Lambda$,
showed that $q_0$ must be positive (equation (6) of
\citet{Deb1999}), whereas inclusion of $\Lambda$ has presented us
a situation in which we can suggest that $q$ must be negative.
This indicates the inconsistency of the result of \citet{Deb1999}
and the present work can be regarded as an improvement over that
so far as the present status of the Universe is concerned.

It has been mentioned in the introduction that the present cosmic
acceleration has started only recently (a few Gyr. earlier).
Before this accelerating phase,the Universe was expanding with
deceleration. So, at the turnover stage (from deceleration to
acceleration), the deceleration parameter $q$ must have changed
its sign. Let us try to find out this signature flip of $q$. Now,
equation (8) can be rewritten as
\begin{eqnarray}
q= -\left(1+\frac{\dot H}{H^2}\right).
\end{eqnarray}
If we assume $\Lambda \simeq H^2$ then (25) reduces to
\begin{eqnarray}
q= -\left(1+\frac{\dot\Lambda}{2CH^3}\right)
\end{eqnarray}
where $C$ is a constant. The above equation (26) tells that when
$\dot \Lambda$ is zero or $\Lambda$ is a constant, then the
Universe always accelerates with a constant acceleration. It may
be mentioned here that, by abandoning $\Lambda$, Einstein obtained
an expanding Universe while the same expanding Universe was
obtained by de Sitter for constant $\Lambda$. But, when $\dot
\Lambda/H^3$ is a function of time, then with a proper choice of
the constant $C$, a signature flip of $q$ can be obtained via
equation (26).

\subsection{Calculation of $\omega$}
It is mentioned earlier that, as a simplest case it is useful to
model dark energy cosmology with a constant equation of state
parameter $\omega$ \citep{Kujat2002,Bartelmann2005}. However, some
useful limits on $\omega$ was suggested by SNIa data, $-1.67 <
\omega < -0.62$ \citep{Knop2003} whereas refined values come from
combined SNIa data with CMB anisotropy and galaxy clustering
statistics which is $-1.33 < \omega < -0.79$ \citep{Tegmark2004}.
Therefore, let us calculate the value of $\omega$ as governed by
the value of $q_0$ and $\Omega_{\Lambda_ 0}$ in the present case.
Putting $q_0=-0.53$ and $\Omega_\Lambda=0.7$ in equation (12), we
get $\omega=0.044$. So, if $q_0=-0.53$, then $\omega>0$ and hence
$p>0$. Thus, the present accelerating Universe may re-collapse in
future if the present value of the deceleration parameter is not
greater than $-0.53$. If $q_0=-0.55$, then we get $p=0$ and hence
a dust-filled Universe. It should be also noted here that for
quintessence, vacuum fluid and phantom energy, the rate of
acceleration should be higher. For instance, when $\omega=-0.5$,
$-1.0$ and $-2.0$ (note that for their simulations
\citet{Kuhlen2005} consider a range of parameter space:
$\omega=-0.5$, $-0.75$, $-1.0$, $-1.25$ and $-1.5$) then we get
respectively, $q_0=-0.775$, $-1.0$ and $-1.45$. On the other hand,
for stiff fluid ($\omega=1.0$), $q_0=-0.1$. So, more smaller the
value of $\omega$, higher is the rate of acceleration. This higher
acceleration may produce the so-called Big Rip
\citep{Caldwell2003} or Partial Rip \citep{Stefancic2004} scenario
due to divergence of scale factor. Figure 1 shows that throughout
the evolution of the Universe, the EOS parameter has assumed
negative as well as positive values including the particular value
1 (stiff fluid). Moreover, according to the figure, $\omega$
assumes the value 1 when $\Omega_\Lambda$ lies between 0.6 and
0.7. So, in the present paper, through an indirect approach, it
has been possible to arrive at the two interesting physical ideas
of modern cosmology mentioned above. An investigation with a
time-dependent $\omega$ may reveal more interesting features.

\section{Field Equations with Variable Speed of Light}
The field equations with varying $c$ are given by
\begin{eqnarray}
3H^2 = 8\pi G \rho+\Lambda c^2 -3\frac{kc^2}{a^2},
\end{eqnarray}
\begin{eqnarray}
2\frac{a \ddot a}{a^2}+\left(\frac {\dot a}{a}\right)^2+\frac
{kc^2}{a^2}-\Lambda c^2 = - \frac {8\pi Gp}{c^2}.
\end{eqnarray}

For flat Universe, $k=0$ and hence, equations (2) and (3)
respectively reduce to
\begin{eqnarray}
3H^2 = 8\pi G \rho+\Lambda c^2,
\end{eqnarray}
\begin{eqnarray}
2(\dot H+H^2)+H^2-\Lambda c^2 = - \frac {8\pi Gp}{c^2}.
\end{eqnarray}

Equation of state is
\begin{eqnarray}
p=\omega\rho c^2,
\end{eqnarray}
where $\omega$ is the barotropic index.

Let us use the {\it ansatz}
\begin{eqnarray}
\Lambda = \alpha H^2,
\end{eqnarray}
where $\alpha$ is a parameter.

Then, by using (31) and (32) in (30), we get
\begin{eqnarray}
2\dot H + 3H^2 - \alpha c^2H^2 = - 8\pi G \omega\rho.
\end{eqnarray}
From (29) we get
\begin{eqnarray}
(3-\alpha c^2)H^2 = 8\pi G \rho.
\end{eqnarray}
By using (34) in (33), we obtain
\begin{eqnarray}
2\dot H = -(3-\alpha c^2)(1+\omega)H^2
\end{eqnarray}
Suppose $c \propto H^{-1}$, then
\begin{eqnarray}
c = \frac{\beta}{H},
\end{eqnarray}
where $\beta$ is a constant of variation. Here we have assumed a
time-dependence of velocity of light quanta photon such that
$c=c(t) \propto t$. This {\it ad hoc} assumption helps us to solve
the differential equation (35) easily and also yields interesting
cosmological scenarios as can be seen later on.

Hence the equation (35) becomes
\begin{eqnarray}
2\dot H = \frac{1}{H^2}(\alpha \beta^2 - 3H^2)(1+\omega)H^2.
\end{eqnarray}
For pressureless dust, $\omega=0$ and hence equation (37) reduces
to
\begin{eqnarray}
2\dot H =  (\alpha \beta^2 - 3H^2).
\end{eqnarray}
By solving equation (38), we get our solution set as
\begin{eqnarray}
a(t) = a_0\left[\cosh
\sqrt{\frac{3\alpha\beta^2}{4}}t\right]^{\frac{2}{3}},
\end{eqnarray}
$a_0 $ is integration constant. It can be readily observed that at
cosmological time $t=0$ the scale factor becomes $a(t)=a_0$. This
means that the present phenomenological $\Lambda$-dark energy
model is singularity free.

The other solutions for the different physical parameters are
\begin{eqnarray}
H(t)= \sqrt{\frac{\alpha\beta^2}{3}}\left[\tanh
\sqrt{\frac{3\alpha\beta^2}{4}}t\right],
\end{eqnarray}

\begin{eqnarray}
\Lambda(t) = \frac{\alpha^2\beta^2}{3}\left[\tanh^2
\sqrt{\frac{3\alpha\beta^2}{4}}t\right],
\end{eqnarray}

\begin{eqnarray}
c(t) = \sqrt{\frac{3}{\alpha}}\left[\coth
\sqrt{\frac{3\alpha\beta^2}{4}}t\right],
\end{eqnarray}

\begin{eqnarray}
\Omega_m = 1-\left[\coth
\sqrt{\frac{3\alpha\beta^2}{4}}t\right]^2,
\end{eqnarray}
\begin{eqnarray}
\Omega_{\Lambda} = \frac{\alpha}{3}.
\end{eqnarray}

\section{Physical Features of the Model}
As can be seen, from the above solutions, we have
\begin{eqnarray}
\Omega_m + \Omega_{\Lambda} = 1 + \frac{\alpha}{3} - \left[\coth
\sqrt{\frac{3\alpha\beta^2}{4}}t\right]^2.
\end{eqnarray}
Then it is easy to see that as $t$ tends to infinity, the sum of
matter and dark-energy densities approaches $\alpha/3$. For the
present Universe, $\Omega_m+\Omega_{\Lambda} \simeq 1$ and hence
$\alpha \simeq 3$.

\begin{figure}
\begin{center}
\vspace{0.5cm}
\includegraphics[width=0.4\textwidth]{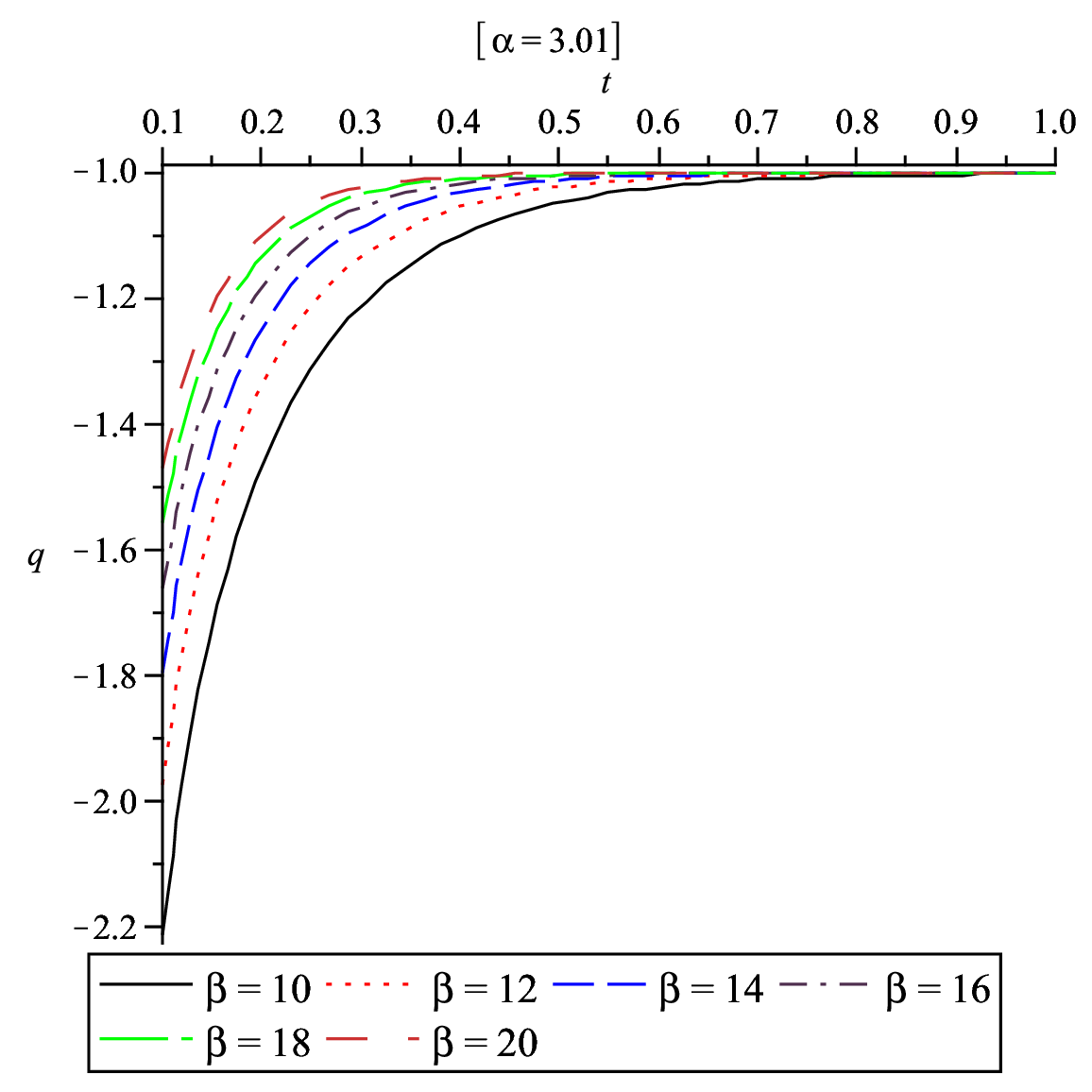}
\caption{The variation of the deceleration parameter q with
respect to $t$ for various values of $\beta$ and fixed $\alpha
=3.01$.}
\end{center}
\end{figure}

\begin{figure}
\begin{center}
\vspace{0.5cm}
\includegraphics[width=0.4\textwidth]{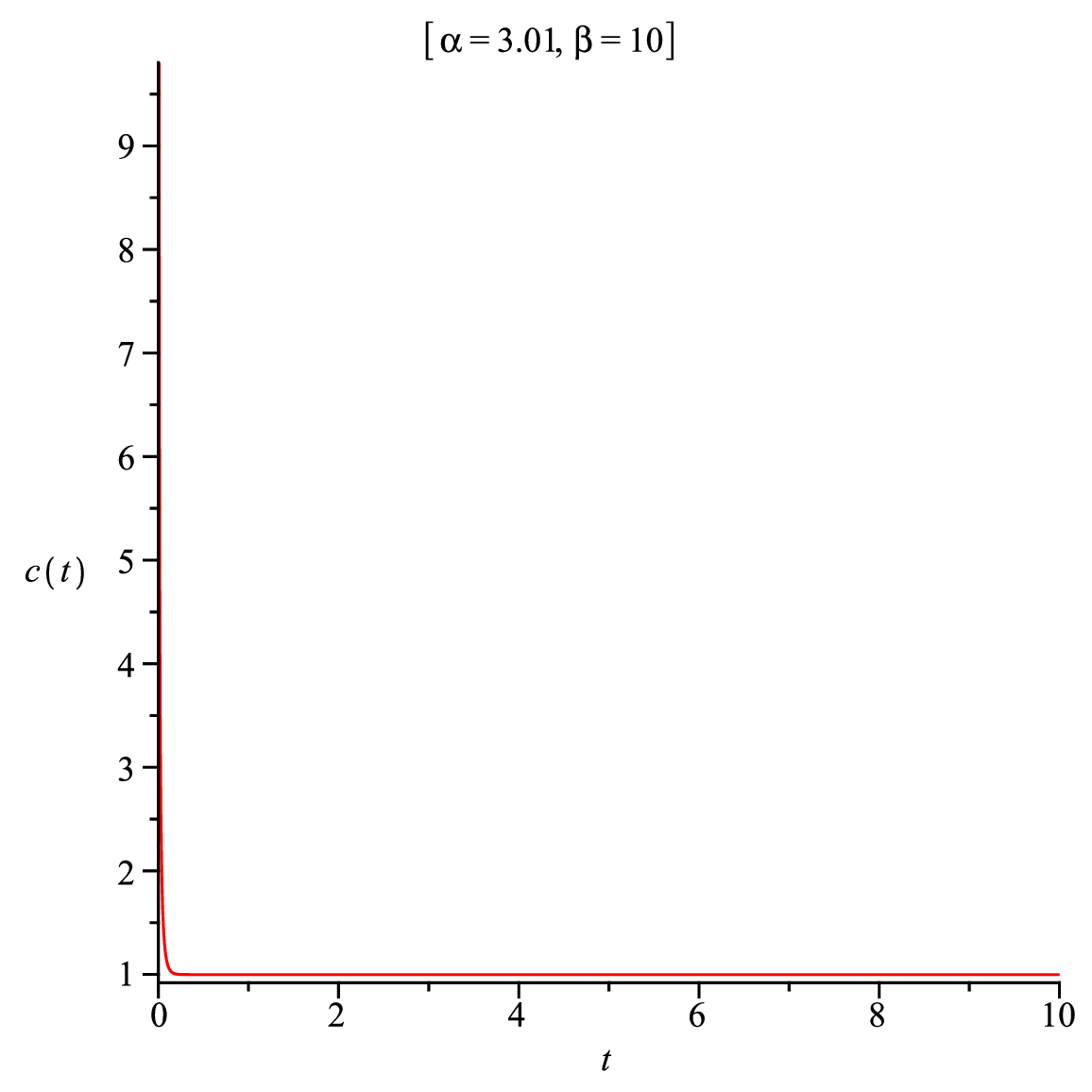}
\caption{The variation of the velocity of light $c(t)$ with
respect to $t$ for various values of $\beta$ and fixed $\alpha
=3.01$.}
\end{center}
\end{figure}

Also, the value of the deceleration parameter $q$ is given by
\begin{eqnarray}
q = -\left[1+ \frac {3}{2 \sinh \sqrt{\frac{3\alpha\beta^2}{4}}t}
\right].
\end{eqnarray}

For physical validity, both $\alpha$ and $\beta$ must be
non-negative. Moreover, equation (41) shows that $\Lambda$ is
always positive irrespective of the values of the parameters
$\alpha$ and $\beta$. This result is consistent with the present
accelerating Universe where a positive $\Lambda$ acts as a
repulsive force to generate the acceleration. Since the quantity
within the bracket in the expression for $q$ is clearly positive,
so $q$ is always negative. Thus, we are getting an
ever-accelerating Universe without any signature flipping of $q$
unlike the modern accepted model in which the Universe was
decelerating in the past and is accelerating at present (Fig. 2).
But, that signature flipping has been obtained already for a
number of variable $\Lambda$ models of phenomenological character
with constant $c$~\citep{Ray2009,Mukhopadhyay2010}. It can be
observed from the Fig. 3 that at the early stage of the evolution
of the Universe, the velocity of light was greater than the
present constant value. However, at the late stages, the value
become exactly the present accepted value. However, this
affirmation is valid only if Eq. (36) can be written in the
following form: $\beta = c_0 H_0$, where $c_0$ would be the
present speed of light. This implies that variation in velocity of
light is not permitted for phenomenological variable $\Lambda$
models as also reported elsewhere by \citet{Ghosh2012}.

\section{Conclusions}
The cosmological term $\Lambda$ and the velocity of light $c$ are
two important quantities in the field equations of Einstein. In
the present work, their specific roles have been studied for an
Universe where dark energy and dark matter are two major
constituents. Though variability of $\Lambda$ is favoured by many
workers for avoiding the cosmological constant problem,
coincidence problem etc., yet the case of constant $\Lambda$
cannot be entirely ruled out. Hence, in the first part of the
paper, the importance of inclusion of the $\Lambda$ term in the
field equations has been demonstrated in the context of the
present accelerating Universe without resorting to any particular
$\Lambda$ model with special reference to the work of
\citet{Deb1999}.

In the next part of the present work, by choosing a well known
time-dependent $\Lambda$ model, viz. $\Lambda \propto H^2$, it has
been shown that the chosen model does not permit the variability
of the speed of light. Moreover, it has been already mentioned in
the Introduction that the equivalence of the chosen model with
other two phenomenological $\Lambda$ models have been shown by
\citet{Ray2007a}. This means that those two models also will not
permit any kind of change in the speed of light. Apart from
arriving at some specific conclusions regarding $\Lambda$ and $c$,
the case of signature flipping of the deceleration parameter $q$
and the crucial role of the equation of state parameter $\omega$
have also been discussed. These are two other salient features of
the present work.

\acknowledgments One of the authors (SR) would like to express his
gratitude to the authority of IUCAA, Pune for providing him the
Associateship Programme under which a part of this work was
carried out.

\end{document}